\documentclass[aps,pre,twocolumn,a4paper,10pt,notitlepage,footinbib,superscriptaddress,longbibliography]{revtex4-1}
\usepackage[english]{babel}
\usepackage{lmodern}
\usepackage[latin9]{inputenc}
\usepackage{endnotes}
\usepackage{amssymb,amsmath,amsfonts}
\usepackage{textcomp}
\usepackage{braket}
\usepackage{soul}
\usepackage{graphicx}% Include figure files
\usepackage{dcolumn}% Align table columns on decimal point
\usepackage{bm}% bold math
\usepackage{xcolor}
\usepackage{soul}

\begin{document}

\title{From non-equilibrium Green's functions 
to Lattice Wigner: A toy model for quantum nanofluidics simulations}
\author{S. Succi}
\affiliation{Center for Life Nano Science@La Sapienza, Istituto Italiano di Tecnologia, 00161 Rome, Italy}
\affiliation{Istituto per le Applicazioni del Calcolo CNR, via dei Taurini 19, 00185 Rome, Italy}
\author{M. Lauricella}
\affiliation{Istituto per le Applicazioni del Calcolo CNR, via dei Taurini 19, 00185 Rome, Italy} 
\author{A. Tiribocchi}
 \affiliation{Istituto per le Applicazioni del Calcolo CNR, via dei Taurini 19, 00185 Rome, Italy}
 \affiliation{INFN ``Tor Vergata'' Via della Ricerca Scientifica 1, 00133 Rome, Italy
}

\begin{abstract}
Recent experiments of fluid transport in nano-channels have shown evidence of a coupling between 
charge-fluctuations in polar fluids  and electronic excitations in graphene solids,
which may lead to  a significant reduction of friction 
a phenomenon dubbed "negative quantum friction". In this paper, we present a semi-classical mesoscale Boltzmann-Wigner lattice kinetic model of quantum-nanoscale transport and perform a numerical study of the effects of the quantum interactions on the evolution of a one-dimensional nano-fluid subject to a periodic external potential. It is shown that the effects of quantum fluctuations become visible once the quantum length scale (Fermi wavelength) of the quasiparticles becomes comparable to the lengthscale of the external potential. Under such conditions, quantum fluctuations are mostly felt on the odd kinetic moments, while the even ones remain nearly unaffected  because they are "protected" by thermal fluctuations. It is hoped that the present Boltzmann-Wigner lattice model and extensions thereof may
offer a useful tool for the computer simulation of quantum-nanofluidic transport phenomena at scales of engineering relevance. 
\end{abstract}

\maketitle

\section{Introduction}

In the last three decades the Lattice Boltzmann (LB) method
has offered a powerful bridge between the atomistic and
macroscopic description of flowing matter, with a broad
spectrum of applications across many regimes and scales
of motion \cite{succi2018lattice,benzi_phys_rep}.
Although the LB formalism extends to the quantum \cite{SUCCI1993327} and relativistic \cite{GABBANA20201}
realms, its overwhelming body of application is focussed on
classical physics, most notably complex fluids and soft matter \cite{physrep}.

However, the relentless progress of nanotechnology is
exposing a growing set of problems whereby quantum 
phenomena need to be explicitly accounted for, a paradigmatic
example being the water flow in carbon nanotubes \cite{gao2017nanofluidics,yu_2023,lizee1,lizee2,bui}.
Recently, it has been surmised that quantum interfacial
effects in dipolar fluids \cite{bocquet1,bocquet2}
and ionic fluids \cite{coquiPNAS} may contribute a sizeable reduction
of the drag experienced by water molecules in the proximity
of graphene confining walls, a phenomenon called negative quantum friction which  could be crucial for the design of surfaces with low hydrodynamic friction.
Such quantum interfacial effects should in principle be treated
by ab-initio quantum statistical mechanics methods, such as
the non-equilibrium Green's function (NEGF) techniques. 
However, due to its steep computational cost, NEGF  
is usually replaced by quantum extensions of 
molecular dynamics \cite{wang2009molecular}. Even so, reaching to
spatial and especially temporal scales of experimental 
relevance remains a major challenge.
There is therefore scope for further coarse-graining,
a task at which LB methods have proved very 
efficient, especially for soft flowing matter applications.  

In this paper, we develop a mathematical framework taking from
NEGF to LB, and most notably to high-order LB schemes \cite{karlinprl,sbragaglia_pre,karlin_pre,ansumali_scirep} capable
of capturing the interplay between classical and quantum non-equilibrium
fluctuations, which lies at the heart of quantum nano-fluidic transport, including
negative quantum friction effects.
The main advantage of a numerical approach based on the LB method lies in its capability of reaching space-time scales up two-three times larger than the ones of typical (quantum) molecular
dynamics simulations \cite{frenkel}, thus easing the disclosure of new physics emerging at the mesoscale.
More specifically, a high order one-dimensional LB method
with third order quantum forcing terms 
is used to model the evolution of a nano-fluid in the
presence of an external periodic potential. 
Our results suggest that, if the length scale
of the quantum force is comparable with that of the external potential, 
quantum fluctuations are found to disturb 
odd moments (i.e. current and energy flux) of the distribution functions,
whereas such moments are screend from quantum effects if the length scale of the potential is larger.
On the contrary, even moments are generally  shielded by thermal fluctuations, which prevail over quantum ones.

The paper is organized as follows. In section \ref{negf1} we shortly recap the NEGF formalism and its
link to the Wigner equation. Sections \ref{negf_lb} and \ref{ho_lb} are dedicated to discussing the derivation of
a high order LB method from the Wigner equation, while section \ref{hyd_curr} highlights the application of the method to
the realistic case of a hydronic current drive \cite{bocquet1,bocquet2}. Finally, in section \ref{d1q5} we describe
the implementation of the one-dimensional LB model with quantum forces and in section \ref{num_res} we present the numerical results,
where we study the effect of such forces on the evolution of the power moments of the distribution functions subject to a periodic potential.
The main findings and conclusions are summarized in the final section.

\section{The non-equilibrium Green's function}\label{negf1}

Following Refs. \cite{kadanoff,rammer}, we start from a quantum many-body 
system described by the quantum wavefunction operator  $\Psi(x,t)$. 
The NEGF formalism is based on the 
Green's function associated with the particle 
generation and destruction operators
$\hat \Psi$ and $\hat \Psi^+$ 
\begin{equation}
G(1,2) = <\hat \Psi(1) \hat \Psi^+(2)>,
\end{equation}
where $1$ and $2$ denote two distinct positions in four-dimensional
spacetime $({\bf x},t)$ 
and brackets denote ensemble averaging over a set of quantum configurations.

By setting ${\bf x}=({\bf x}_1+{\bf x}_2)/2$, 
${\bf r}={\bf x}_1-{\bf x}_2$, $t=(t_1+t_2)/2$ and $s= t_1-t_2$
and taking the Fourier-transform, we obtain the 
associated Wigner function describing the distribution of
{\it quasiparticles} in eight-dimensional
phase-spacetime $({\bf x},{\bf p},t,E)$ \cite{wig_physrep}
\begin{equation}
W({\bf x},t;{\bf p},E) = \int e^{-i({\bf p}\cdot {\bf r}-Es)/\hbar} G({\bf x},{\bf r};t,s) d{\bf r} ds. 
\end{equation}

By assuming weak interactions, which means that the quasi-particles obey the one-valued dispersion relation $E=E({\bf p})$, and integrating upon the energy variable,  the Wigner equation read as follows
\begin{equation}
\partial_t W  + {\bf v}\cdot \nabla_x W +
\Theta ({\bf F}\cdot \nabla_p W) = C,
\end{equation}

where ${\bf v}={\bf p}/m$ (being $m$ the mass of the quasiparticle and ${\bf p}$ its momentum) and $C$ is a collision term resulting from
the scattering processes between the quasiparticles.
In the above $\Theta$ denotes a non-local functional
in energy-momentum space resulting from quantum 
interference effects.
In explicit form
\begin{equation}
\Theta=\sum_{k\in N_{odd}} \left(\frac{\hbar}{2i}\right)^{|k|-1} \frac{1}{k!} F_k \partial_p^k,
\end{equation}
where $F_k = -\partial_x^k U$ \cite{lat_wig}, $U$ being the one-body effective potential.
The Wigner function bears a close resemblance to a classical 
probability distribution function, in that its kinetic
moments can be associated to the quasiparticle density and current,
in close analogy with classical hydrodynamics. 
This property is key to establish a consistent bridge 
with the Boltzmann equation.
However, its quantum nature is reflected by the fact 
that $W$ is a pseudo-probability distribution which can take both signs as a result of quantum interference \cite{wig_physrep}.

Mathematically, this is due to the higher order derivatives
in momentum space, which probe higher order spatial derivatives
of the potential.
Since these derivatives in the streaming term
scale like odd powers of the quantum Knudsen number
\begin{equation}\label{qKn}
q= \lambda_F/\delta,
\end{equation}
quantum interference effects are responsible for the non-positivity of the Wigner function. Eq.(\ref{qKn}) is the analogue of the Knudsen number $Kn=\lambda/\delta$, where the molecular mean free path $\lambda$ is replaced by the Fermi wavelength
$\lambda_{F}=\frac{\hbar}{m v_{F}}$ 
(being $\hbar$ the reduced Planck constant and $v_F$ the Fermi speed) and $\delta$ is the typical lengthscale. 
Note that for quadratic potentials, the Wigner function recovers 
positive-definiteness (because the quantum force is identically zero) and becomes fully classical, hence quantum effects are exposed by third order spatial derivatives onward.

\section{From NEGF to Boltzmann and  high-order Lattice Boltzmann}\label{negf_lb}

For the homogeneous case, close to equilibrium,
the dependence on ${\bf x}$ and $t$ of the Wigner function drops out.
 However, since we shall be dealing with quantum 
non-equilibrium  transport phenomena, such an assumption
is not justified. 
A Boltzmann-like equation can be derived under two major assumptions.
First, the heterogeneity must be weak at the quantum scale, which is determined by 
the Fermi wavelength $\lambda_F$. Formally
\begin{equation}
q \ll 1,
\end{equation}  
which means that at the transport scale (set by $\delta$),
the quantum excitations (quasiparticles) are 
localized, hence they can be treated as quasi classical particles.

The second assumption is that quantum excitations should be weakly
interacting, meaning that their self-energy must be small
as compared to classical kinetic energy $k_BT$ (where $k_B$ is the Boltzmann constant and $T$ is the temperature).
Formally,
\begin{equation}
Fr = \frac{F \delta}{k_BT} \ll 1,
\end{equation}  
where $Fr$ is the Froude number and $F=-\nabla U$.
The weak-interaction regime $Fr \ll 1$ permits to associate
a single-valued dispersion relation to the quantum excitations,
i.e. $\omega=\omega(k)$ and $\gamma=\gamma(k)$, where
$k=p/\hbar$ is the wavenumber and $\omega$ and $\gamma$ are
the real and imaginary part of the complex wave-frequency.
% (energy).
The former controls phase changes (propagation) and the latter
amplitude changes (decay/stability). 
Under such condition the Wigner distribution can be expressed
in the so called in-shell representation 
\begin{equation}
W({\bf x},{\bf p},t,E) = f({\bf x},{\bf p},t) \delta [E-E({\bf p})],
\end{equation}
so that the energy-dependence can be integrated out to
yield a Boltzmann equation in six-dimensional 
phase-space plus time
\begin{equation}
\label{BEQ}
\partial_t f  + {\bf v}\cdot \nabla_x f + {\bf F}\cdot \nabla_p f = C(f,f),
\end{equation}
where $C(f,f)$ is a semiclassical collision operator.
In the sequel it proves expedient to replace $C$ with the corresponding
single-relaxation time expression \cite{bgk} 
\begin{equation}
C = \frac{f-f^{eq}}{\tau},
\end{equation}
where $f^{eq}$ is a Bose-Einstein or Fermi-Dirac local
equilibrium for bosons and fermions and $\tau$ is the relaxation time.

With the Boltzmann equation at hand, the route to LB follows the standard 
protocol, with the important proviso that high-order lattices (HOL) are no luxury, but play a vital role instead.
To this purpose let us reminds that in the theory of classical
LB, HOL are usually employed to go beyond the hydrodynamic regime
and describe strong non-equilibrium effects associated with
non-negligible Knudsen numbers, i.e $Kn \gg 1$.

For quantum nanofluidics, there are two additional motivations:
first, quantum local equilibria demand energy conservation, hence
they need to be formulated on lattices extending beyond the first
Brillouin region \cite{coelho2014lattice}. Second, as discussed earlier on,
in the presence of quantum interference, higher order derivatives
in momentum space need to be accounted for, which again commends the resort to HOL.   

\section{Quantum interference and High-Order LB}\label{ho_lb}

The former aspect is discussed in full detail in \cite{coelho2014lattice}, hence
in the following we focus on the latter.
Let us consider, for example, the third order term in one spatial dimension 
for simplicity, i.e. $F_3(x) \partial_{p}^3 f$ where $p=p_x$.
With reference to a generic microscopic property 
$\phi(p)$, the change per unit time of the macroscopic moment
$\Phi_3(x) = \int \phi(p) f(x,p) dp$ due to the third order force is given by 
$$
\dot \Phi_3(x) = F_3(x) \int \phi(p) \partial_{p}^3 f(x,p) dp. 
$$  
On the assumption that all boundary contributions 
vanish at infinity in momentum spaces, repeated 
integration by parts delivers
\begin{equation}
\dot \Phi_3(x) = F_3(x) \int f(x,p) \partial_{p}^3 \phi(p) dp, 
\end{equation}
which gives zero for moments below third order.
However, microscopic quantities of order three (i.e. the skewness)
couple to the zero-th order moment, which is the fluid density.
If, for instance, $\phi(p) = p^3/6$,  we obtain $\dot \Phi_3(x) = F_3(x) n(x)$;
likewise, $\phi(p)=p^4/24$ contributes $\dot \Phi_3(x) = F_3(x) J$
and so on. This shows a long-range coupling in momentum space
as a result of quantum interference, whence the need of high-order
lattices.
A detailed list of 2D lattices (whose implementation is challenging but conceptually straightforward) 
with up to sixteenth order isotropy can be found in \cite{sbragaglia2007generalized}.

Indeed, previous numerical simulations have shown that the use
of HOL leads to  more accurate results in the case
of anharmonic (fourth-order) potentials, confirming that kinetic
moments of order above three do couple to the hydrodynamic
sector \cite{solorzano2018lattice,brewer2016lattice}.
This is because third order derivatives in momentum space, as applied to
an Hermite mode of order $n$, excites modes of
order $n+3$ in the Hermite ladder.

\section{Prospective application to hydronic current drive}\label{hyd_curr}

In this section we discuss the relevant regimes for
hydronic current drive nanodevices \cite{succi2024keldysh}.
To convey a concrete idea of a typical application scenario, let us 
consider a fluid of water molecules flowing in a nano-channel, 
say a carbon nanotube, confined by 
carbon walls, either graphite or graphene \cite{bocquet1,bocquet2}. 

Water is driven by an external pressure gradient and 
dissipates energy and momentum on the solid walls. 
However, at variance with the classical picture, whereby such 
dissipation is due to classical interaction of the 
water molecules with solid molecules at the wall,
new interfacial interactions need to be considered.
In particular, the nanoscale fluctuations of the 
water molecules give rise to corresponding nanoscale fluctuations of the molecular 
charge (dubbed {\it hydrons}), which couple to electronic degrees of freedom in the 
solid wall via screened Coulomb interactions.  
At the same time, classical mechanical collisions
of the water molecules with the solid walls generate phonon 
excitations in the solid. 
Due to phonon-electron scattering, these two mechanisms  induce a 
net motion of both excitations, namely a "phonon wind" and an "electron wind", which
are ultimately stabilized by momentum and energy dissipation on 
the solid crystal, thus closing the energy balance.

Ab-initio analysis based on the quantum non-equilibrium Keldysh formalism
predicts that the interaction between 
water and hydrons in the liquid, and electrons and phonons in the solid,
leads to a broad variety of energy exchange patterns 
between the flowing water and the electron-phonon "fluids" in the solid, including
the possibility that the electrons may return 
energy and momentum back to the liquid, thereby leading to a reduction
of the friction experienced by water, a mechanism  dubbed 
"negative quantum friction" \cite{bocquet1,bocquet2}.
\begin{figure}
\centering
\includegraphics[width=0.5\textwidth]{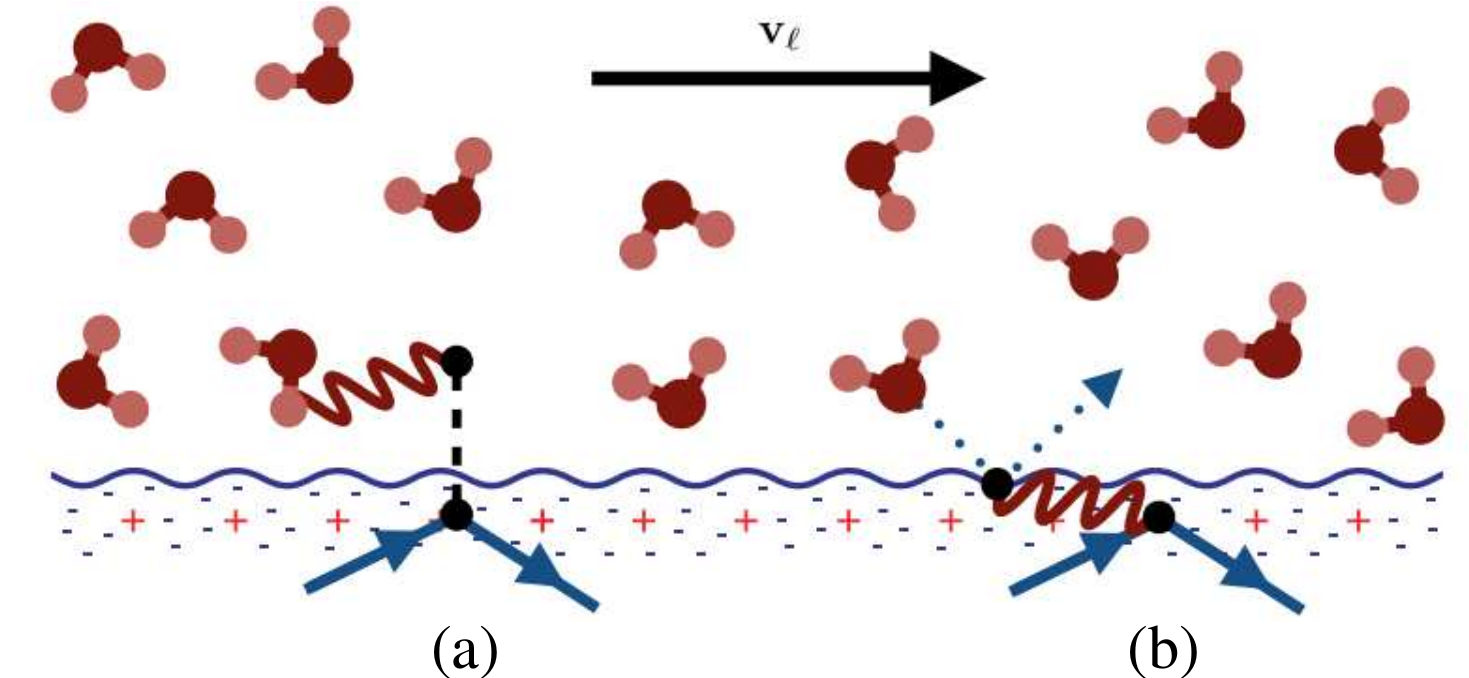}
\caption{Two different mechanisms driving the electronic current: (a) water molecules subject to charged fluctuations (hydrons, red wavy line) transferring momentum to the electrons (black dots) in the solid through Coulomb interaction; (b) phonons (wavy line), excited by water molecules collisions, drive electrons in the solid. The figure is adapted from Ref.\cite{bocquet2}.}
\label{qfric}
\end{figure}

Such quantum effects can be estimated in terms of the quantum Knudsen number $q$ (defined in Eq.\ref{qKn}), where the spatial scale of the hydrodynamic fields $\delta$ is assumed
comparable to the lengthscale of the interaction potential.
The quantum Knudsen number controls the strength of the quantum force versus the 
classical one, namely
\begin{equation}
\frac{F_3 \partial_{p}^3 f}{F_1 \partial_p f}\sim q^2,
\end{equation}
where we have taken $\partial_x \sim \delta^{-1}$ and $\partial_p \sim 1/(m v_{F})$.
Another useful dimensionless group is the "quantumness", hereby defined as
the ratio of the Fermi wavelength to a characteristic mean free path
\begin{equation}
    \mathcal{Q} \equiv \lambda_F/\lambda.
\end{equation}
By definition, the $q$ and $\mathcal{Q}$ are related via the classical mean free path as $q = \mathcal{Q} {Kn}$.
This shows that the condition 
$\mathcal{Q} > 1$
indicates that we are dealing with quantum fluids.
To be noted at variance with $q$, which is flow-dependent, that the 
quantumness is inherently a fluid property.
It also displays an upper bound dictated by the 
celebrated AdS-CFT minimum viscosity bound \cite{kovtun,trachenko}, 
which states that any fluid should fulfill the following inequality 
\begin{equation}\label{mvb}
\frac{\eta}{s} \ge \frac{1}{4 \pi} \frac{\hbar}{k_B},
\end{equation}
where $\eta$ is the dynamic viscosity of the fluid and $s$ is the entropy density.
The above inequality is nearly saturated by strongly interacting fluids, such as
quark-gluon plasmas, whereas ordinary fluids lie about two or more orders of magnitude above.
By recasting Eq.(\ref{mvb}) in terms of the "quantumness", we readily obtain
$\mathcal{Q}  \le 4 \pi$, where we have taken the entropy per particle of order unity. 

Next we consider typical values for a quanto-nanofluidic application,
with reference to a nanotube of  diameter $D=10$ nm and length $L=100$ nm, with  solid wall thickness $a=1$ nm. 
The electron Fermi wavelength is
$\lambda_F = h/\sqrt{m E_f} \sim 2.5$ nm,
where 
we have taken $m=m_e/10$ for the effective electron mass in graphene and $E_F\simeq 100$ meV \cite{bocquet1}.
Assuming longitudinal propagation of the electrons and a transport scale $\delta \sim 10$nm, we have $q\sim 0.25$.
This shows that the electron mean free path is comparable with the Fermi
wavelength, hence the electronic excitations can be treated semi-classically.
We note that in our case, the value of the quantumness 
$\mathcal{Q} \sim 1 $ points indeed to a strongly interacting
fluid,  but still consistent with the AdS-CFT bound. 

\section{The D1Q5 model with quantum forces}\label{d1q5}

\begin{figure}[htbp]
\centering
\includegraphics[width=0.4\textwidth]{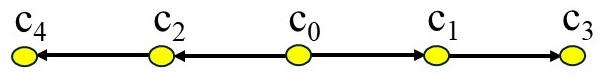}
\caption{D1Q5 lattice Boltzmann scheme. 
Black arrows indicate lattice speeds $c_i$ where $i=0,1,2,3,4$.}
\label{fig0}
\end{figure}

Here we describe a one-dimensional lattice Boltzmann method to study dynamics of the first five moments of the Wigner distribution function.
We consider a D1Q5 lattice consisting of five discrete speeds $c_{i_x}=c_i$, where 
$c_i=\frac{\Delta x}{\Delta t}$ (with $i=0,1,2,3,4$), $\Delta x$ is the lattice step 
and $\Delta t$ is the time step, and modulus $c_0=0$, $c_1=+1$, $c_2=-1$, $c_3=+2$, $c_4=-2$  (see Fig.\ref{fig0}).  
A set of distribution functions $f_i(x,t)$, defined on each site $x$ and time $t$, evolves following a 
discrete Boltzmann equation \cite{succi2018lattice}
\begin{equation}\label{LB_eq}
f_i(x + c_i\Delta t, t +\Delta t)=(1-\omega)f_i(x,t)+\omega f_i^{eq}(x,t)+S_i(x,t),
\end{equation}
where $\omega$ is a frequency tuning the relaxation towards the equilibrium and controlling the fluid viscosity $\nu=(1/\omega -1/2)c_s^2\Delta t$ (with $c_s$ lattice sound speed and $c_s^2=1$), $f_i^{eq}$ are the local equilibrium populations and $S_i$ are the source terms \cite{succi2018lattice}. 
Following common practice in LB theory \cite{succi2018lattice,halim}, the former are computed as a second-order Taylor expansion in the fluid velocity $u$ (with $u=u_x$) at low Mach number
\begin{equation}
f_i^{eq}=w_i\rho\left[1+\frac{uc}{c^2_s}+\frac{u^2(c_ic_i-c^2_s)}{2c_s^4}\right],
\end{equation}
where $\rho$ is the fluid density and $w_i$ is a set of weights with values $w_0=\frac{6}{12}$, $w_1=w_2=\frac{2}{12}$ and $w_3=w_4=\frac{1}{12}$. Also, the fluid density $\rho$ and the fluid momentum $\rho u$ can be computed from the moments of the distributions $f_i$ as $\rho=\sum_if_i$ and $\rho u=\sum_if_ic_i$.

Note that the actual populations $f_i$ can be written in terms of the kinetic moments $M_i$ as follows
\begin{equation}
f_i(x,t) = w_i \sum_{k=0}^4 M_k(x,t) V_i^k,
\end{equation}
where $V_i^k$ is a set of orthogonal  eigenvectors
\begin{eqnarray}
&&V^0_i = 1_i = [1,1,1,1,1],\\
&&V^1_i = c_i = [0,1,-1,2,-2],\\
&&V^2_i = c_i^2-c_s^2 = [-1,0,0,3,3],\\
&&V^3_i = c_i^3-3 c_ic_s^2 = [0,-2,2,2,-2],\\
&&V^4_i = c_i^4-4 c_i^2c_s^2 +c_s^4 = [1,-2,-2,1,1].
\end{eqnarray}

The kinetic moments are thus given by
\begin{equation}
M_k(x,t) = \sum_{i=0}^4 f_i(x,t) V_i^k,
\end{equation}
which are used to systematically derive the equations of motion and the forcing terms.

\subsection{Equations of motion}\label{eq_mot}

By multiplying Eq.(\ref{LB_eq}) by $1,c_i,c_i^2,c_i^3,c_i^4$ and summing
up, the equations of motion take the following form
\begin{eqnarray}
&&\partial_t M_0 + \partial_x M_1 = 0,\label{mom0}\\
&&\partial_t M_1 + \partial_x M_2 = S_1^{cl},\label{mom1}\\
&&\partial_t M_2 + \partial_x M_3 = -\omega(M_2-M_2^{eq})+S_2^{cl},\label{mom2}\\
&&\partial_t M_3 + \partial_x M_4 = -\omega(M_3-M_3^{eq})+S_3^{cl}+S_3^q,\label{mom3}\\
&&\partial_t M_4 + \partial_x M_5 = -\omega(M_4-M_4^{eq})+S_4^{cl}+S_4^q,\label{mom4}
\end{eqnarray}
where $S_i^{cl,q}$ are the classical and quantum forces, whose computation 
is presented in the next subsection.

Rather than studying the physics of the kinetic moments $M_k$, we prefer monitoring the effect of the quantum force on the power moments $P_k$, which are given by
\begin{equation}\label{pw_mom}
P_k = \sum_{i=0}^4 f_i c_i^k,\;\;k=0,1,2,3,4.
\end{equation} 
Indeed, besides carrying a direct physical interpretation, these moments  allow for an easier analysis of the origin of the quantum effects which are expected to play a role in the absence of thermal fluctuations. 
In this respect, one can easily prove that 
$P_0=\rho$, $P_1=\rho (u+\theta_1)$,  $P_2=\rho(u^2 + \theta_2)$, 
$P_3 = \rho(u^3 + u \theta_2 + \theta_3)$
and $P_4 = \rho(u^4 + 6u^2 \theta_2 + \theta_4)$, where 
$\rho\theta_p=\sum_{i=1}^4f_i(c_i-u)^p$, being $\theta_p$ the correlator.

Note that the odd correlators $\theta_1$ and $\theta_3$ 
vanish at equilibrium, while the even correlators do not,
since they carry  the contribution of thermal fluctuations, namely 
$\theta_2^{eq}=c_s^2$ is the square of the thermal speed and $\theta_4^{eq}=3c_s^4$ is the flatness of the equilibrium distribution. These values corresponds to the central moments a Gaussian profile, where $\theta_3^{eq}$ is the skewness and $\theta_4^{eq}$ is the kurtosis.
At equilibrium one has $P_1^{eq}=\rho u$,
$P_2^{eq}=\rho (u^2+c_s^2)$ and $P_3^{eq}=\rho u(u^2+c^2_s)$, corresponding to the fluid current, the energy density and the energy flux density, respectively.
The non-equilibrium components of the correlators are associated with
non-equilibrium fluctuations driven by heterogeneity and they are
responsible for irreversible transport phenomena. 

\subsection{Forcing terms}
Next we consider the effect of the forcing terms.
For classical forces we have
\begin{equation}
S^{cl}(x,p,t) = -F_1(x) \partial_p f, 
\end{equation}
where $F_1(x) = -\partial_x U(x)$ and $U(x)$ is the external potential, while
the associated moments are
\begin{equation}
S_k^{cl}(x,t) = F_1(x) \int H_k(p) \partial_p f dp = - F_1(x) \int f \partial_p H_k(p) dp,
\end{equation}
where $H_k(p)$ is an Hermite basis in continuum velocity space. Simple integration by parts delivers
$S^{cl}_0 = 0$, $S^{cl}_1 = \rho F_1$,
$S^{cl}_2 = 2JF_1$, $S^{cl}_3 = 3JuF_1$,
$S^{cl}_4 = 4Ju^2F_1$, where $J=\rho u$. 
The contributions to the discrete distributions can be cast
in the same form as the discrete distributions themselves, namely
\begin{equation}
S_i(x,t)= w_i \sum_{k=0}^4 S_k(x,t) V_i^k,
\end{equation}
which are the source terms associated with the classical force.
The same procedure applied to the quantum force
\begin{equation}
S^{q}(x,p,t) = -F_3(x) \partial_{p}^3 f dp 
\end{equation}
delivers $S^{q}_0 = 0$, $S^{q}_1 = 0$,
$S^{q}_2 = 0$, $S^{q}_3 = 6\rho F_3$,
$S^{q}_4 = 24JF_3$, 
where $F_3(x) = -\partial_{x}^3U$.
Note that the quantum force does not act directly upon the first
three moments, namely density, current and energy, although it can affect them 
through the gradients of the moments of order three and four.
Also, since $F_3(x)$ stems from a third order derivative in space
of the external potential, the quantum force is most effective on the short scales. 
For a potential lenghtscale $\delta$, the ratio of the quantum force to the classical one
scales exactly like $q^2$.

\section{Numerical results}\label{num_res}

\subsection{Constant force}
As a benchmark test, we consider a one-dimensional fluid subject to a constant force due, for example, to an external electric field {\bf E}, and in the presence of a friction force $-\gamma{\bf u}$, where $\gamma$ is the friction coefficient. This is a basic test to verify the Ohm's law, where 
an electric current $I$ (induced by the electric field) attains, at the steady state, a constant value depending on $\gamma$ (which would play the role of an electric resistance). Note that, since we deal with a 1d fluid, the inclusion of such a friction force is crucial for a correct modeling of dissipative effects, which are not fully captured by the sole shear viscosity.

\begin{figure}
\includegraphics[width=0.5\textwidth]{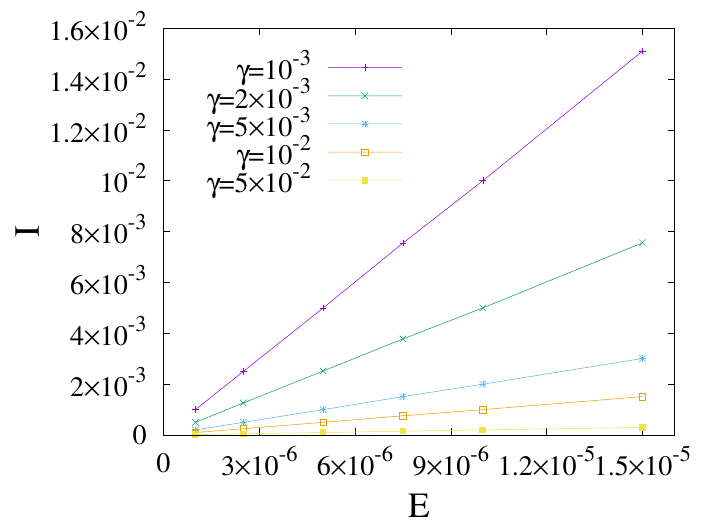}
\caption{This plot shows the behavior of the steady state current $I$ induced by a constant electric field $E$. For increasing values of $\gamma$ the slope decreases,  accurately following the equation $I=\frac{\rho q E}{\gamma}$. In our simulations $q$ is set to one. The values of $\gamma$ can be computed, from a linear fit, as the inverse of the slope.}
\label{fig3_tmp}
\end{figure}

In a one dimensional channel, the current $I$ coincides with its density $J$ (i.e. the first kinetic moment $M_1$) which, by definition, is equal to $\rho u$. At the steady state, the balance between friction and electric force leads to $u=\frac{qE}{m\gamma}$ (being $q$ the charge  of the particle and $m$ the mass), finally yielding $I\propto \gamma^{-1}$. This is basically the behavior shown in Fig.\ref{fig3_tmp}, where  we plot the steady state values of $I$ {\it versus} the magnitude of the electric field $E$. As expected, the values of $I$ diminish for increasing $\gamma$, whose values can be computed as the inverse of the slope of each curve.

Note that quantum fluctuations would be absent in this system, since they are expected to emerge when higher order derivatives of the potential survive. In the next section we consider precisely this scenario.

\subsection{Periodic potential}

To inspect the effect of the quantum fluctuations, we study the relaxation of a one-dimensional fluid which,
in addition to the forces previously considered, is also subject to a periodic one, stemming from an external periodic potential. 

The latter one has the following  form
\begin{equation}
U(x)=-U_0cos(k_nx),
\end{equation}
where $k_n=\frac{\pi}{L}n_w$ and $n_w$ is the wavenumber. 
This leads to $F_1(x)=U_0k_n sin(k_nx)$ and $F_3(x)=- U_0 k_n^3 sin(k_nx)$.
This choice ensures the existence of high order derivatives necessary to model quantum fluctuations, while the presence of a constant force plus a frictional one guarantee a nonzero steady state current and the inclusion of dissipative effects, respectively.
The simulations are initialized with an inhomogeneous density distribution following 
a Gaussian profile and run for 
approximately $10^4$ time steps.
If not stated otherwise, we study this system for two values of $n_w$, i.e. $n_w=8$ (low frequency regime) and 
$n_w=32$ (high frequency regime), two values of $\omega$, i.e. $\omega=0.5$ and $\omega=1$, $E=10^{-6}$, $\gamma=10^{-3}$,
$U_0=10^{-3}$ and 
$L_x=128$ in the absence and presence of the quantum force $F_3$.
Also, the lattice spacings are set to $\Delta x=1$ and $\Delta t=1$ which would approximately correspond to $1$ nm and $1$ ps in real units. This leads to a channel length of roughly $128$ nm  and an experiment lasting for $\sim 10$ ns.
If we take $\lambda_F\sim 5\Delta x$ and $\delta\sim L/n_w$, we have $q\sim 0.3$ for $n_w=8$ and $q\sim 1.25$ for $n_w=32$, thus quantum effects are expected to become visible at high wavenumbers.

\subsection{Low wavenumber regime}
\begin{figure*}
\includegraphics[width=1.0\textwidth]{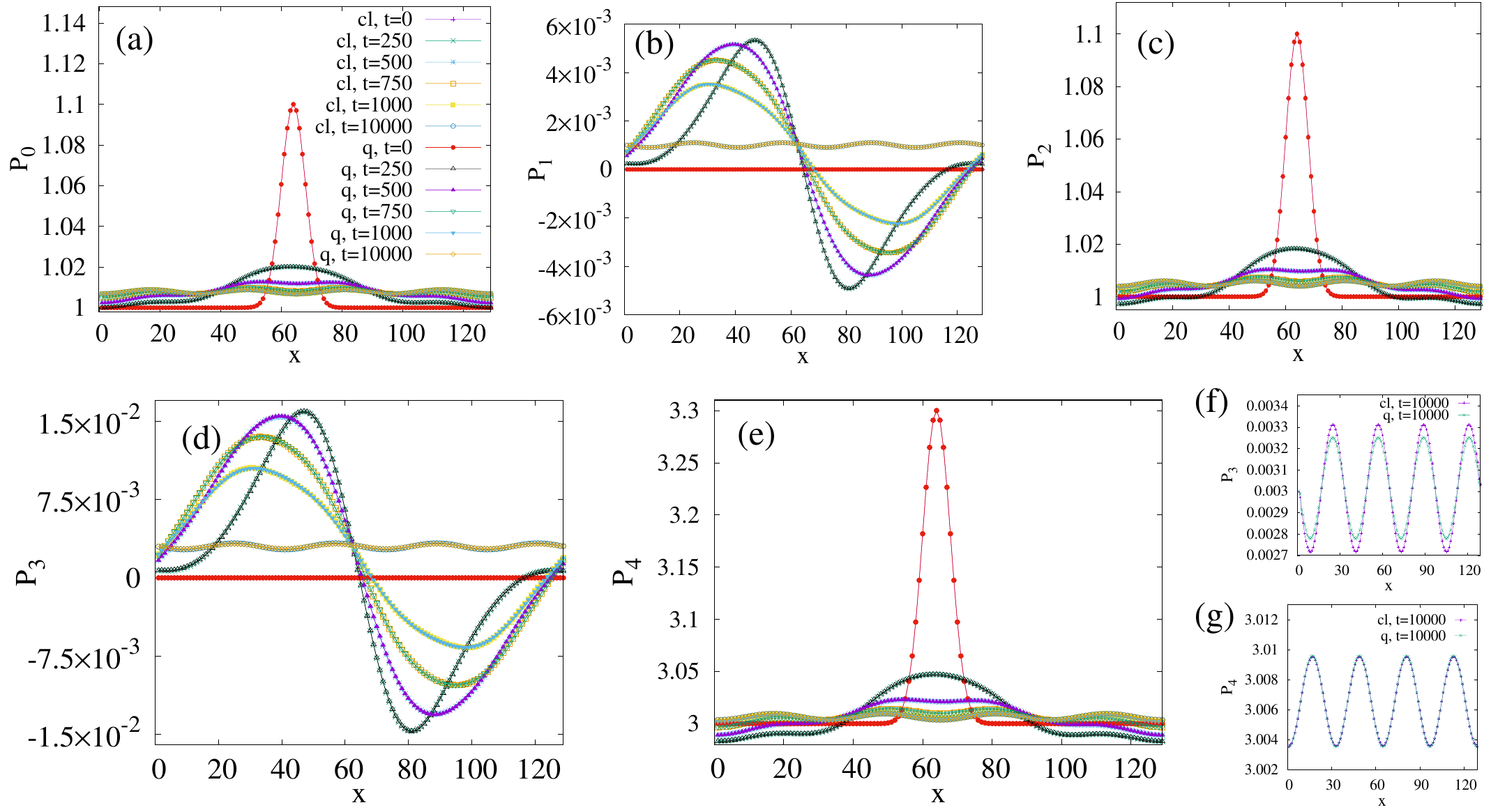}
\caption{Power moments $P_0$, $P_1$, $P_2$, $P_3$ and $P_4$
at simulation time $t=0$, $250$, $500$, $750$, $1000$, $10000$ for $n_w=8$ and $\omega=1$. 
The classical ("cl") and quantum ("q") distributions of the first three moments (a,b,c) are basically indistinguishable. While $P_0$, $P_2$ and $P_4$ relax in a similar manner starting from a Gaussian profile,  $P_1$ 
exhibits an intermediate wave-like behavior, which 
progressively turns into a sinusoidal profile fluctuating around $10^{-3}$, i.e. the steady state value of the current.
The effect of the quantum force is slightly visible for $P_3$ 
where the quantum distribution barely deviates from the classical one, 
as highlighted in the late-time profile shown in (f). Such effects are essentially indiscernible for $P_4$ (e and g).
}\label{fig3}
\end{figure*}

In Fig.\ref{fig3} we show the time evolution of the five power moments $P_k$ of classical and quantum 
distributions for $n_w=8$ and $\omega=1$ (setting the numerical viscosity to $\nu=0.5$), where the 
initial Gaussian profile of the density is centered at $L/2$ with a standard deviation $\sigma=4$. 
Classical profiles are obtained by setting $F_3=0$, while quantum ones include $F_3$.
At $t=0$, all moments except $P_0$ are zero.
Our results show that the first three moments, $P_0$, $P_1$ and $P_2$, are not affected by quantum forces to 
any appreciable extent, not even through gradients of higher order moments. 
Both  classical and quantum distributions of $P_0$ and $P_2$ gradually relax towards an almost 
flat profile with values slightly larger than $1$ (Fig.\ref{fig3}a,c),  while $P_1$
displays a wave-like symmetric profile (Fig.\ref{fig3},b).
The first moment $P_1$ is positive for $x<L/2$ and negative elsewhere,  with fixed zeroes at the boundaries 
and  at $L/2$ (i.e. where density gradients are constant), while maximum and minimum 
(corresponding to the inflection points of the density profile) gradually shift towards 
lower values, until the current vanishes everywhere. 

Quantum effects are found to very mildly affect only the moment $P_3$
(see Fig.\ref{fig3}d,f), whose quantum distribution slightly deviates from the classical one, which displays a wave-like symmetric profile overall akin to $P_1$. 
The distribution $P_4$ follows the typical behavior of the even moments and is basically unaffected by any quantum deviations (Fig.\ref{fig3}e,g). 
The different response of the moments $P_3$ and $P_4$
to the quantum force depends on 
the fact that the effect of such a force for the even moments is masked by the thermal fluctuations which, on the contrary, vanish at equilibrium (i.e. when $u\simeq 0$) for the odd moments (see the explicit expressions of the power moments in Sec.\ref{eq_mot}). 
Note also that amplitude and frequency of both distributions at late times (Fig.\ref{fig3}e,g) remain essentially consistent with the values of amplitude $U_0$  and wavenumber $n_w$ set by the periodic potential $U(x)$. 

Deviations from the classical distribution can be approximately quantified in terms of the percentage error 
$\Delta_k=100\times \frac{|P_k^{cl}-P_k^{q}|}{P_k^{cl}}$, where $P_k^{cl}$ and $P_k^{q}$ stand for classical and quantum distributions. As previously mentioned, $\Delta_k$ is negligible for all moments except $P_3$, where the highest value is found around $1\%$.

These results point towards a picture where, as long as $q$ remains below one (i.e. when $n_w$ is relatively low), the effect of the quantum force on the  moments is basically negligible. In the next section we show that this scenario changes significantly at increasing values of $n_w$.

\subsection{High wavenumber regime} 
\begin{figure*}
\includegraphics[width=1.0\textwidth]{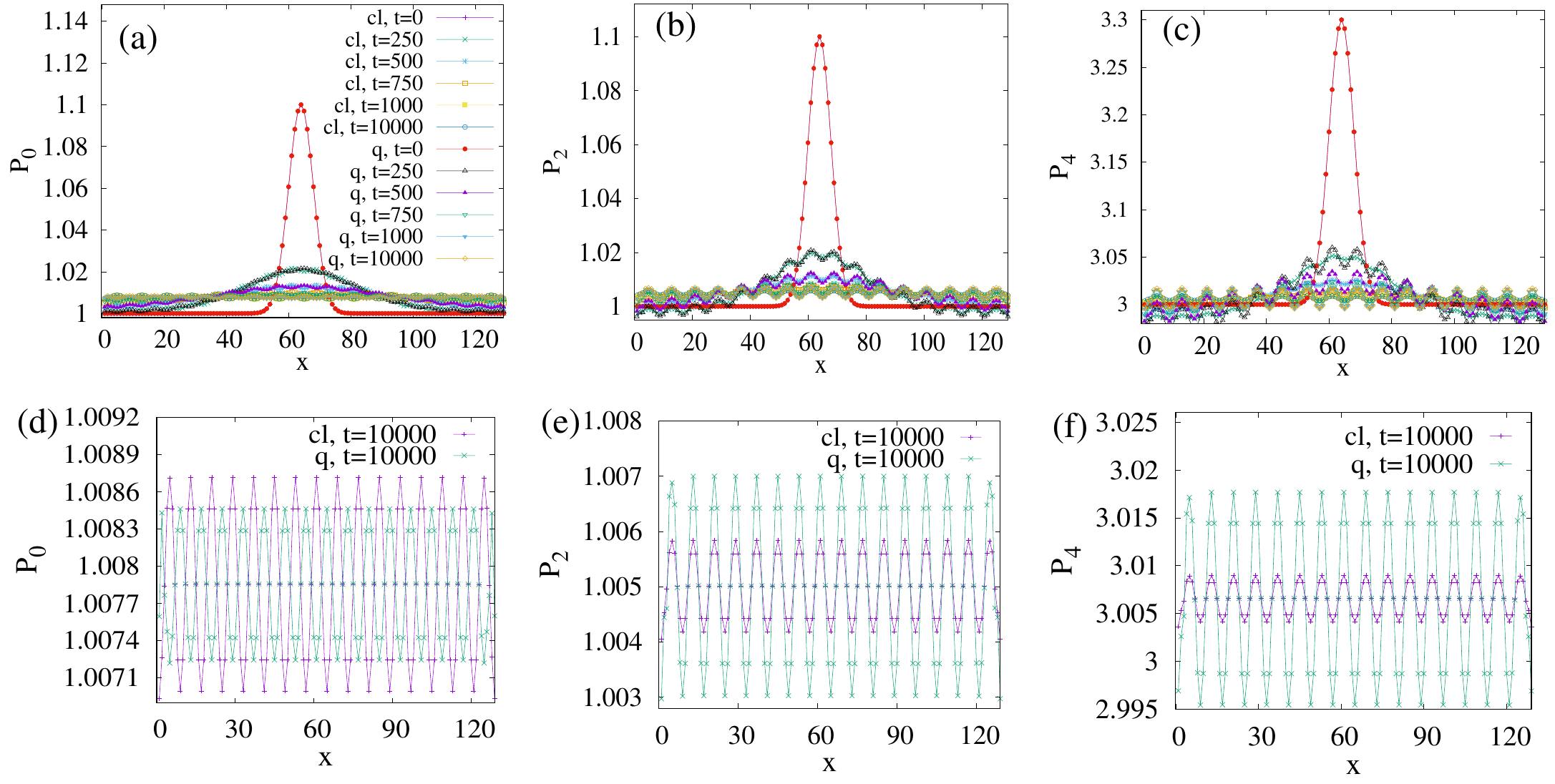}
\caption{Even power moments 
$P_0$ (a), $P_2$ (b), $P_4$ (c) at $t=0$, $250$, $500$, $750$, $1000$, $10000$ for $n_w=32$ and $\omega=1$. Figures (d), (e), (f), highlight the moment profiles at $t=10000$. 
The time evolution is overall akin to that observed for $n_w=8$. However, here the quantum 
force generally induces an  amplitude change, either larger or smaller than that of the classical profile, as shown in the late-time profiles (d,e,f). The misalignment of maxima and minima between the distributions (for example in $P_3$ and $P_4$) results from the signed amplitude of the forces, $F_1\propto U_0k_n$ and $F_3\propto -U_0k^3_n$.}
\label{fig4}
\end{figure*}

\begin{figure*}
\includegraphics[width=0.7\textwidth]{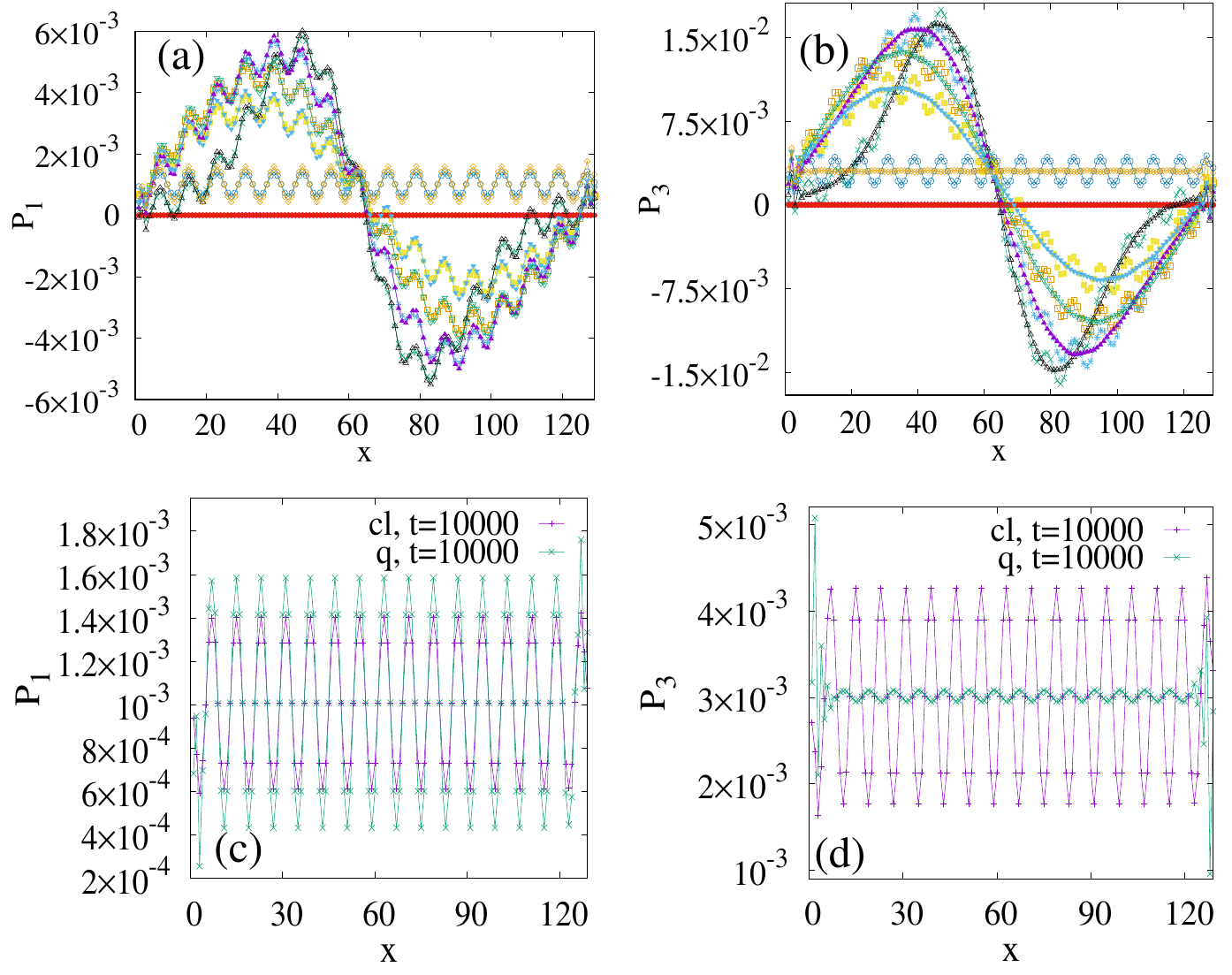}
\caption{Odd power moments 
$P_1$ (a), $P_3$ (b), at $t=0$, $250$, $500$, $750$, $1000$, $10000$ for $n_w=32$ and $\omega=1$. Figures (d), (e), (f), highlight the moment profiles at $t=10000$. 
The time evolution of the distributions is similar to the one observed for lower wavenumbers, although here larger values of $n_w$ induce  wider undulations. 
The quantum force yields, once again, a change of amplitude with respect to the classical signal (c,d).}
\label{fig5}
\end{figure*}

In Fig.\ref{fig4} and Fig.\ref{fig5} we show, for example, the time evolution of the even and odd power moments for $n_w=32$ and $\omega=1$. 
While the time behavior (Fig.\ref{fig4}a,b,c and Fig.\ref{fig5}a,b) is overall akin to the one observed  for lower values of $n_w$ (except that here a higher wavenumber induces larger undulations), 
the effect of the quantum force is clearly visible on the profiles of {\it all} power moments (as shown in the late time configurations of Fig.\ref{fig4}d,e,f and Fig.\ref{fig5}c,d).  
Indeed, although  the quantum force enters explicitly only the  moments $M_3$ and $M_4$ in Eqs.(\ref{mom3}-\ref{mom4}) (and thus $P_3$ and $P_4$), its effect actually conditions lower moment through spatial derivatives (for instance $\partial_x M_3$ in Eq.(\ref{mom2})) resulting from the hierarchical structure of the equations of motion.

In the present problem, the quantum force manifests through a change of amplitude of the distributions, which display a wave-like behavior with a well-defined frequency (set by $n_w$).
Considering, for example, the late-time profiles of $P_2$ and $P_4$ in Fig.\ref{fig4}e-f, the quantum force slightly amplifies the classical signal, while the opposite holds for $P_0$, where the
minima are found to correspond to the maxima of the quantum distribution, leading to an apparent phase-shift effect. This actually happens because $F_1$ and $F_3$ carry opposite signs and different amplitudes, where $F_1\propto k_nU_0$ and $F_3\propto -U_0k^3_n$.
Odd moments generally show similar features, although 
the amplitude of the quantum signal can considerably change, either increase, as in $P_1$, or substantially decrease as in $P_3$ (see Fig.\ref{fig5}c,d). This is caused, once again, by the absence of thermal fluctuations at equilibrium for the odd moments, thus exposing such moments to quantum effects.
In this respect, the odd moments show values of $\Delta_k$ considerably higher than the even ones. Specifically, we find that the highest values are $\Delta_1\simeq 13\%$ and $\Delta_3\simeq 30\%$ (a difference arising
essentially because the quantum force is mitigated at lower moments), while $\Delta_{0,2,4}$ are lower than $1\%$, although not negligible. 

\begin{figure*}
\includegraphics[width=1.0\textwidth]{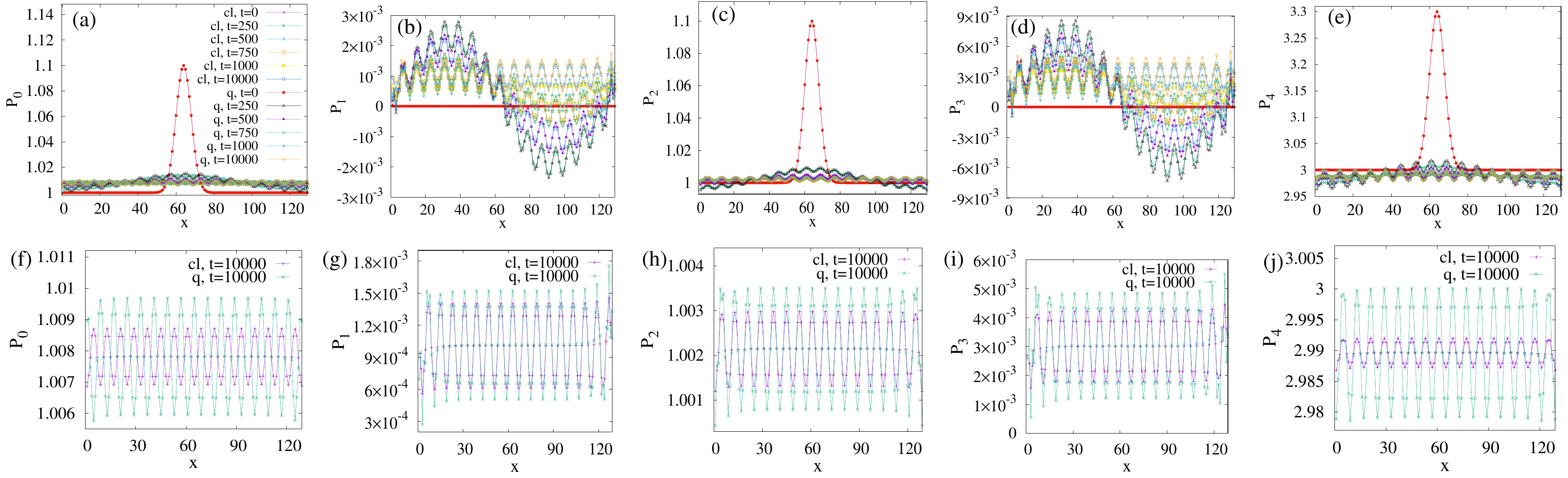}
\caption{Power moments 
$P_0$ (a), $P_1$ (b), $P_2$ (c), $P_3$ (d) and $P_4$ (e) at $t=0$, $250$, $500$, $750$, $1000$, $10000$ for $n_w=32$ and $\omega=0.5$. Figures (f), (g), (h), (i) and (j) highlight the moment profiles at $t=10000$. 
Unlike the previous case, here lower values of $\omega$ mainly affect odd moments, which display an undulated profile with smaller amplitude. At late times, both $P_1$ and $P_3$ do not show appreciable modulations; this is because, at larger viscosity, the fluid equilibrates on a shorter time scale.}
\label{fig6}
\end{figure*}

It is also of interest to understand whether modifying the shear viscosity alters such a picture. In Fig.\ref{fig6} we show the time evolution of the power moments for $\omega=0.5$, which sets a numerical viscosity $\eta= 1.5$. Note that, although $0<\omega<2$, the AdS-CFT minimum viscosity bound would impose a lower upper bound, which would be safely set at $\omega\simeq 1.5$.
Once again quantum effects condition all moments, in a way overall akin to the scenario observed for the larger values of $\omega$ and with similar values of $\Delta_k$. 
Note that increasing the viscosity leads to a slight decrease (both in the classical and quantum distributions) of the amplitudes of the odd moments $P_1$ ad $P_3$ (corresponding to  current and energy flux respectively) with respect to $\omega=1$,
essentially because larger values of $\eta$ entail a higher dissipation.

\section{Conclusions}

Summarizing, we have presented a mathematical derivation of a high-order Boltzmann-Wigner lattice kinetic 
equation starting from non-equilibrium Green's function
formulation of quantum non-equilibrium transport phenomena.
Simulations of a minimal D1Q5 lattice with third order quantum forcing terms
for the case of a periodic potential indicate that in the semiclassical regime
($q<1$), the lowest hydrodynamic modes are well protected against  quantum interference 
effects as long as the wavenumber $n_w$ (controlling the  characteristic lenghtscale  $\delta$) is sufficiently low.
In actual practice, it is reasonable to assume that the length scale
of the effective potential be significantly larger than the relevant
Fermi wavelength, namely $k_F \delta > 1$, where $k_F$ is the modulus of the Fermi 
wavevector. 
Of course, such an assumption needs to be checked on a case-by-case
basis, but the fact remains that the lowest order moments (i.e. density
and current) can only be affected by quantum interference effects on
condition of strong coupling  with classical non-equilibrium effects carried by the spatial gradient 
of the "handshaking" moment $P_2$. This is indeed  observed at larger values of $n_w$ (when $q>1$), where the presence of the quantum force generally yields a substantial change of amplitude of the classical signal. 
This is particularly relevant for odd moments where thermal fluctuations vanish at equilibrium,  thus allowing quantum effects to emerge.
It is therefore plausible to expect that the lattice Boltzmann-Wigner equation discussed in this paper may provide an efficient description of a variety of quantum-nanofluidic phenomena, although further studies are needed to investigate the interplay between external fields, dissipation and fluctuation-induced potentials. In this respect, the present work may have a limited relevance to experiments because of the use of toy potentials and the restriction to one-dimensional fluids. We plan to partially overcome these drawbacks in  future works, which will be concerned with the study of two dimensional fluids (simulated on high order lattices) in the presence of more realistic potentials, such as the screened Coulomb one, and density fluctuations of the liquid.

%\section*{Data availability statement}
%The data that support the findings of this study are available from the corresponding author upon reasonable request.

%\section*{Conflict of interest}
%The authors have no conflicts to disclose.

\begin{acknowledgments}

We thank L. Bocquet of \'Ecole Normale Sup\'erieure, N. Kavokine of EPFL and E. Kaxiras of Harvard Physics Department for many valuable hints and discussions. M. L. and A. T. thank the Italian National Group for Mathematical Physics of INdAM (GNFM-INdAM) for the support.

\end{acknowledgments}

\nocite{*}
\bibliography{aipsamp}% Produces the bibliography via BibTeX.

\providecommand{\noopsort}[1]{}\providecommand{\singleletter}[1]{#1}%
\begin{thebibliography}{32}%
\makeatletter
\providecommand \@ifxundefined [1]{%
 \@ifx{#1\undefined}
}%
\providecommand \@ifnum [1]{%
 \ifnum #1\expandafter \@firstoftwo
 \else \expandafter \@secondoftwo
 \fi
}%
\providecommand \@ifx [1]{%
 \ifx #1\expandafter \@firstoftwo
 \else \expandafter \@secondoftwo
 \fi
}%
\providecommand \natexlab [1]{#1}%
\providecommand \enquote  [1]{``#1''}%
\providecommand \bibnamefont  [1]{#1}%
\providecommand \bibfnamefont [1]{#1}%
\providecommand \citenamefont [1]{#1}%
\providecommand \href@noop [0]{\@secondoftwo}%
\providecommand \href [0]{\begingroup \@sanitize@url \@href}%
\providecommand \@href[1]{\@@startlink{#1}\@@href}%
\providecommand \@@href[1]{\endgroup#1\@@endlink}%
\providecommand \@sanitize@url [0]{\catcode `\\12\catcode `\$12\catcode
  `\&12\catcode `\#12\catcode `\^12\catcode `\_12\catcode `\%12\relax}%
\providecommand \@@startlink[1]{}%
\providecommand \@@endlink[0]{}%
\providecommand \url  [0]{\begingroup\@sanitize@url \@url }%
\providecommand \@url [1]{\endgroup\@href {#1}{\urlprefix }}%
\providecommand \urlprefix  [0]{URL }%
\providecommand \Eprint [0]{\href }%
\providecommand \doibase [0]{http://dx.doi.org/}%
\providecommand \selectlanguage [0]{\@gobble}%
\providecommand \bibinfo  [0]{\@secondoftwo}%
\providecommand \bibfield  [0]{\@secondoftwo}%
\providecommand \translation [1]{[#1]}%
\providecommand \BibitemOpen [0]{}%
\providecommand \bibitemStop [0]{}%
\providecommand \bibitemNoStop [0]{.\EOS\space}%
\providecommand \EOS [0]{\spacefactor3000\relax}%
\providecommand \BibitemShut  [1]{\csname bibitem#1\endcsname}%
\let\auto@bib@innerbib\@empty
%</preamble>
\bibitem [{\citenamefont {Succi}(2018)}]{succi2018lattice}%
  \BibitemOpen
  \bibfield  {author} {\bibinfo {author} {\bibfnamefont {S.}~\bibnamefont
  {Succi}},\ }\href@noop {} {\emph {\bibinfo {title} {The lattice Boltzmann
  equation: for complex states of flowing matter}}}\ (\bibinfo  {publisher}
  {Oxford University Press},\ \bibinfo {year} {2018})\BibitemShut {NoStop}%
\bibitem [{\citenamefont {Benzi}\ \emph {et~al.}(1992)\citenamefont {Benzi},
  \citenamefont {Succi},\ and\ \citenamefont {Vergassola}}]{benzi_phys_rep}%
  \BibitemOpen
  \bibfield  {author} {\bibinfo {author} {\bibfnamefont {R.}~\bibnamefont
  {Benzi}}, \bibinfo {author} {\bibfnamefont {S.}~\bibnamefont {Succi}}, \ and\
  \bibinfo {author} {\bibfnamefont {M.}~\bibnamefont {Vergassola}},\ }\bibfield
   {title} {\enquote {\bibinfo {title} {The lattice {B}oltzmann equation:
  theory and applications},}\ }\href@noop {} {\bibfield  {journal} {\bibinfo
  {journal} {Phys. Rep.}\ }\textbf {\bibinfo {volume} {222}},\ \bibinfo {pages}
  {145--197} (\bibinfo {year} {1992})}\BibitemShut {NoStop}%
\bibitem [{\citenamefont {Succi}\ and\ \citenamefont
  {Benzi}(1993)}]{SUCCI1993327}%
  \BibitemOpen
  \bibfield  {author} {\bibinfo {author} {\bibfnamefont {S.}~\bibnamefont
  {Succi}}\ and\ \bibinfo {author} {\bibfnamefont {R.}~\bibnamefont {Benzi}},\
  }\bibfield  {title} {\enquote {\bibinfo {title} {Lattice {B}oltzmann equation
  for quantum mechanics},}\ }\href@noop {} {\bibfield  {journal} {\bibinfo
  {journal} {Physica D: Nonlinear Phenomena}\ }\textbf {\bibinfo {volume}
  {69}},\ \bibinfo {pages} {327--332} (\bibinfo {year} {1993})}\BibitemShut
  {NoStop}%
\bibitem [{\citenamefont {Gabbana}\ \emph {et~al.}(2020)\citenamefont
  {Gabbana}, \citenamefont {Simeoni}, \citenamefont {Succi},\ and\
  \citenamefont {Tripiccione}}]{GABBANA20201}%
  \BibitemOpen
  \bibfield  {author} {\bibinfo {author} {\bibfnamefont {A.}~\bibnamefont
  {Gabbana}}, \bibinfo {author} {\bibfnamefont {D.}~\bibnamefont {Simeoni}},
  \bibinfo {author} {\bibfnamefont {S.}~\bibnamefont {Succi}}, \ and\ \bibinfo
  {author} {\bibfnamefont {R.}~\bibnamefont {Tripiccione}},\ }\bibfield
  {title} {\enquote {\bibinfo {title} {Relativistic lattice {B}oltzmann
  methods: {T}heory and applications},}\ }\href@noop {} {\bibfield  {journal}
  {\bibinfo  {journal} {Phys. Rep.}\ }\textbf {\bibinfo {volume} {863}},\
  \bibinfo {pages} {1--63} (\bibinfo {year} {2020})}\BibitemShut {NoStop}%
\bibitem [{\citenamefont {Tiribocchi}\ \emph {et~al.}(2025)\citenamefont
  {Tiribocchi}, \citenamefont {Durve}, \citenamefont {Lauricella},
  \citenamefont {Montessori}, \citenamefont {Tucny},\ and\ \citenamefont
  {Succi}}]{physrep}%
  \BibitemOpen
  \bibfield  {author} {\bibinfo {author} {\bibfnamefont {A.}~\bibnamefont
  {Tiribocchi}}, \bibinfo {author} {\bibfnamefont {M.}~\bibnamefont {Durve}},
  \bibinfo {author} {\bibfnamefont {M.}~\bibnamefont {Lauricella}}, \bibinfo
  {author} {\bibfnamefont {A.}~\bibnamefont {Montessori}}, \bibinfo {author}
  {\bibfnamefont {J.~M.}\ \bibnamefont {Tucny}}, \ and\ \bibinfo {author}
  {\bibfnamefont {S.}~\bibnamefont {Succi}},\ }\bibfield  {title} {\enquote
  {\bibinfo {title} {Lattice {B}oltzmann simulations for soft flowing
  matter},}\ }\href@noop {} {\bibfield  {journal} {\bibinfo  {journal} {Phys.
  Rep.}\ }\textbf {\bibinfo {volume} {1105}},\ \bibinfo {pages} {1--52}
  (\bibinfo {year} {2025})}\BibitemShut {NoStop}%
\bibitem [{\citenamefont {Gao}\ \emph {et~al.}(2017)\citenamefont {Gao},
  \citenamefont {Feng}, \citenamefont {Guo},\ and\ \citenamefont
  {Jiang}}]{gao2017nanofluidics}%
  \BibitemOpen
  \bibfield  {author} {\bibinfo {author} {\bibfnamefont {J.}~\bibnamefont
  {Gao}}, \bibinfo {author} {\bibfnamefont {Y.}~\bibnamefont {Feng}}, \bibinfo
  {author} {\bibfnamefont {W.}~\bibnamefont {Guo}}, \ and\ \bibinfo {author}
  {\bibfnamefont {L.}~\bibnamefont {Jiang}},\ }\bibfield  {title} {\enquote
  {\bibinfo {title} {Nanofluidics in two-dimensional layered materials:
  inspirations from nature},}\ }\href@noop {} {\bibfield  {journal} {\bibinfo
  {journal} {Chem. Soc. Rev.}\ }\textbf {\bibinfo {volume} {46}},\ \bibinfo
  {pages} {5400--5424} (\bibinfo {year} {2017})}\BibitemShut {NoStop}%
\bibitem [{\citenamefont {Yu}\ \emph {et~al.}(2023)\citenamefont {Yu},
  \citenamefont {Principi},\ and\ \citenamefont {Tielrooij}}]{yu_2023}%
  \BibitemOpen
  \bibfield  {author} {\bibinfo {author} {\bibfnamefont {X.}~\bibnamefont
  {Yu}}, \bibinfo {author} {\bibfnamefont {A.}~\bibnamefont {Principi}}, \ and\
  \bibinfo {author} {\bibfnamefont {K.~J. et~al.}\ \bibnamefont {Tielrooij}},\
  }\bibfield  {title} {\enquote {\bibinfo {title} {Electron cooling in graphene
  enhanced by plasmon-hydron resonance},}\ }\href@noop {} {\bibfield  {journal}
  {\bibinfo  {journal} {Nat. Nanotechnol.}\ }\textbf {\bibinfo {volume} {18}},\
  \bibinfo {pages} {898--904} (\bibinfo {year} {2023})}\BibitemShut {NoStop}%
\bibitem [{\citenamefont {Liz\'ee}\ \emph {et~al.}(2023)\citenamefont
  {Liz\'ee}, \citenamefont {Marcotte}, \citenamefont {Coquinot}, \citenamefont
  {Kavokine}, \citenamefont {Sobnath}, \citenamefont {Barraud}, \citenamefont
  {Bhardwaj}, \citenamefont {Radha}, \citenamefont {Nigu\`es}, \citenamefont
  {Bocquet},\ and\ \citenamefont {Siria}}]{lizee1}%
  \BibitemOpen
  \bibfield  {author} {\bibinfo {author} {\bibfnamefont {M.}~\bibnamefont
  {Liz\'ee}}, \bibinfo {author} {\bibfnamefont {A.}~\bibnamefont {Marcotte}},
  \bibinfo {author} {\bibfnamefont {B.}~\bibnamefont {Coquinot}}, \bibinfo
  {author} {\bibfnamefont {N.}~\bibnamefont {Kavokine}}, \bibinfo {author}
  {\bibfnamefont {K.}~\bibnamefont {Sobnath}}, \bibinfo {author} {\bibfnamefont
  {C.}~\bibnamefont {Barraud}}, \bibinfo {author} {\bibfnamefont
  {A.}~\bibnamefont {Bhardwaj}}, \bibinfo {author} {\bibfnamefont
  {B.}~\bibnamefont {Radha}}, \bibinfo {author} {\bibfnamefont
  {A.}~\bibnamefont {Nigu\`es}}, \bibinfo {author} {\bibfnamefont
  {L.}~\bibnamefont {Bocquet}}, \ and\ \bibinfo {author} {\bibfnamefont
  {A.}~\bibnamefont {Siria}},\ }\bibfield  {title} {\enquote {\bibinfo {title}
  {Strong electronic winds blowing under liquid flows on carbon surfaces},}\
  }\href@noop {} {\bibfield  {journal} {\bibinfo  {journal} {Phys. Rev. X}\
  }\textbf {\bibinfo {volume} {13}},\ \bibinfo {pages} {011020} (\bibinfo
  {year} {2023})}\BibitemShut {NoStop}%
\bibitem [{\citenamefont {Liz\'ee}\ \emph {et~al.}(2024)\citenamefont
  {Liz\'ee}, \citenamefont {Coquinot},\ and\ \citenamefont
  {Mariette}}]{lizee2}%
  \BibitemOpen
  \bibfield  {author} {\bibinfo {author} {\bibfnamefont {M.}~\bibnamefont
  {Liz\'ee}}, \bibinfo {author} {\bibfnamefont {B.}~\bibnamefont {Coquinot}}, \
  and\ \bibinfo {author} {\bibfnamefont {G.~et~al.}\ \bibnamefont {Mariette}},\
  }\bibfield  {title} {\enquote {\bibinfo {title} {Anomalous friction of
  supercooled glycerol on mica},}\ }\href@noop {} {\bibfield  {journal}
  {\bibinfo  {journal} {Nat. Commun.}\ }\textbf {\bibinfo {volume} {15}},\
  \bibinfo {pages} {6129} (\bibinfo {year} {2024})}\BibitemShut {NoStop}%
\bibitem [{\citenamefont {Bui}\ \emph {et~al.}(2023)\citenamefont {Bui},
  \citenamefont {Thiemann}, \citenamefont {Michaelides},\ and\ \citenamefont
  {Cox}}]{bui}%
  \BibitemOpen
  \bibfield  {author} {\bibinfo {author} {\bibfnamefont {A.~T.}\ \bibnamefont
  {Bui}}, \bibinfo {author} {\bibfnamefont {F.~L.}\ \bibnamefont {Thiemann}},
  \bibinfo {author} {\bibfnamefont {A.}~\bibnamefont {Michaelides}}, \ and\
  \bibinfo {author} {\bibfnamefont {S.~J.}\ \bibnamefont {Cox}},\ }\bibfield
  {title} {\enquote {\bibinfo {title} {Classical quantum friction at
  water-carbon interfaces},}\ }\href@noop {} {\bibfield  {journal} {\bibinfo
  {journal} {Nano Lett.}\ }\textbf {\bibinfo {volume} {23}},\ \bibinfo {pages}
  {580--587} (\bibinfo {year} {2023})}\BibitemShut {NoStop}%
\bibitem [{\citenamefont {Kavokine}\ \emph {et~al.}(2022)\citenamefont
  {Kavokine}, \citenamefont {Bocquet},\ and\ \citenamefont
  {Bocquet}}]{bocquet1}%
  \BibitemOpen
  \bibfield  {author} {\bibinfo {author} {\bibfnamefont {N.}~\bibnamefont
  {Kavokine}}, \bibinfo {author} {\bibfnamefont {M.~L.}\ \bibnamefont
  {Bocquet}}, \ and\ \bibinfo {author} {\bibfnamefont {L.}~\bibnamefont
  {Bocquet}},\ }\bibfield  {title} {\enquote {\bibinfo {title}
  {Fluctuation-induced quantum friction in nanoscale water flows},}\
  }\href@noop {} {\bibfield  {journal} {\bibinfo  {journal} {Nature}\ }\textbf
  {\bibinfo {volume} {602}},\ \bibinfo {pages} {84--90} (\bibinfo {year}
  {2022})}\BibitemShut {NoStop}%
\bibitem [{\citenamefont {Coquinot}\ \emph {et~al.}(2023)\citenamefont
  {Coquinot}, \citenamefont {Bocquet},\ and\ \citenamefont
  {Kavokine}}]{bocquet2}%
  \BibitemOpen
  \bibfield  {author} {\bibinfo {author} {\bibfnamefont {B.}~\bibnamefont
  {Coquinot}}, \bibinfo {author} {\bibfnamefont {L.}~\bibnamefont {Bocquet}}, \
  and\ \bibinfo {author} {\bibfnamefont {N.}~\bibnamefont {Kavokine}},\
  }\bibfield  {title} {\enquote {\bibinfo {title} {Quantum feedback at the
  solid-liquid interface: {F}low-induced electronic current and its negative
  contribution to friction},}\ }\href@noop {} {\bibfield  {journal} {\bibinfo
  {journal} {Phys. Rev. X}\ }\textbf {\bibinfo {volume} {13}},\ \bibinfo
  {pages} {011019} (\bibinfo {year} {2023})}\BibitemShut {NoStop}%
\bibitem [{\citenamefont {Coquinot}\ \emph {et~al.}(2024)\citenamefont
  {Coquinot}, \citenamefont {Bocquet},\ and\ \citenamefont
  {Kavokine}}]{coquiPNAS}%
  \BibitemOpen
  \bibfield  {author} {\bibinfo {author} {\bibfnamefont {B.}~\bibnamefont
  {Coquinot}}, \bibinfo {author} {\bibfnamefont {L.}~\bibnamefont {Bocquet}}, \
  and\ \bibinfo {author} {\bibfnamefont {N.}~\bibnamefont {Kavokine}},\
  }\bibfield  {title} {\enquote {\bibinfo {title} {Hydroelectric energy
  conversion of waste flows through hydroelectronic drag},}\ }\href@noop {}
  {\bibfield  {journal} {\bibinfo  {journal} {Proceedings of the National
  Academy of Sciences}\ }\textbf {\bibinfo {volume} {121}},\ \bibinfo {pages}
  {e2411613121} (\bibinfo {year} {2024})}\BibitemShut {NoStop}%
\bibitem [{\citenamefont {Wang}\ \emph {et~al.}(2009)\citenamefont {Wang},
  \citenamefont {Ni},\ and\ \citenamefont {Jiang}}]{wang2009molecular}%
  \BibitemOpen
  \bibfield  {author} {\bibinfo {author} {\bibfnamefont {J.-S.}\ \bibnamefont
  {Wang}}, \bibinfo {author} {\bibfnamefont {X.}~\bibnamefont {Ni}}, \ and\
  \bibinfo {author} {\bibfnamefont {J.-W.}\ \bibnamefont {Jiang}},\ }\bibfield
  {title} {\enquote {\bibinfo {title} {Molecular dynamics with quantum heat
  baths: Application to nanoribbons and nanotubes},}\ }\href@noop {} {\bibfield
   {journal} {\bibinfo  {journal} {Phys. Rev. B}\ }\textbf {\bibinfo {volume}
  {80}},\ \bibinfo {pages} {224302} (\bibinfo {year} {2009})}\BibitemShut
  {NoStop}%
\bibitem [{\citenamefont {Chikatamarla}\ and\ \citenamefont
  {Karlin}(2006)}]{karlinprl}%
  \BibitemOpen
  \bibfield  {author} {\bibinfo {author} {\bibfnamefont {S.~S.}\ \bibnamefont
  {Chikatamarla}}\ and\ \bibinfo {author} {\bibfnamefont {I.~V.}\ \bibnamefont
  {Karlin}},\ }\bibfield  {title} {\enquote {\bibinfo {title} {Entropy and
  {G}alilean invariance of lattice {B}oltzmann theories},}\ }\href@noop {}
  {\bibfield  {journal} {\bibinfo  {journal} {Phys. Rev. Lett.}\ }\textbf
  {\bibinfo {volume} {97}},\ \bibinfo {pages} {190601} (\bibinfo {year}
  {2006})}\BibitemShut {NoStop}%
\bibitem [{\citenamefont {Sbragaglia}\ \emph
  {et~al.}(2007{\natexlab{a}})\citenamefont {Sbragaglia}, \citenamefont
  {Benzi}, \citenamefont {Biferale}, \citenamefont {Succi}, \citenamefont
  {Sugiyama},\ and\ \citenamefont {Toschi}}]{sbragaglia_pre}%
  \BibitemOpen
  \bibfield  {author} {\bibinfo {author} {\bibfnamefont {M.}~\bibnamefont
  {Sbragaglia}}, \bibinfo {author} {\bibfnamefont {R.}~\bibnamefont {Benzi}},
  \bibinfo {author} {\bibfnamefont {L.}~\bibnamefont {Biferale}}, \bibinfo
  {author} {\bibfnamefont {S.}~\bibnamefont {Succi}}, \bibinfo {author}
  {\bibfnamefont {K.}~\bibnamefont {Sugiyama}}, \ and\ \bibinfo {author}
  {\bibfnamefont {F.}~\bibnamefont {Toschi}},\ }\bibfield  {title} {\enquote
  {\bibinfo {title} {Generalized lattice {B}oltzmann method with multirange
  pseudopotential},}\ }\href@noop {} {\bibfield  {journal} {\bibinfo  {journal}
  {Phys. Rev. E}\ }\textbf {\bibinfo {volume} {75}},\ \bibinfo {pages} {026702}
  (\bibinfo {year} {2007}{\natexlab{a}})}\BibitemShut {NoStop}%
\bibitem [{\citenamefont {Chikatamarla}\ and\ \citenamefont
  {Karlin}(2009)}]{karlin_pre}%
  \BibitemOpen
  \bibfield  {author} {\bibinfo {author} {\bibfnamefont {S.~S.}\ \bibnamefont
  {Chikatamarla}}\ and\ \bibinfo {author} {\bibfnamefont {I.~V.}\ \bibnamefont
  {Karlin}},\ }\bibfield  {title} {\enquote {\bibinfo {title} {Lattices for the
  lattice {B}oltzmann method},}\ }\href@noop {} {\bibfield  {journal} {\bibinfo
   {journal} {Phys. Rev. E}\ }\textbf {\bibinfo {volume} {79}},\ \bibinfo
  {pages} {046701} (\bibinfo {year} {2009})}\BibitemShut {NoStop}%
\bibitem [{\citenamefont {Namburi}\ \emph {et~al.}(2016)\citenamefont
  {Namburi}, \citenamefont {Krithivasan},\ and\ \citenamefont
  {Ansumali}}]{ansumali_scirep}%
  \BibitemOpen
  \bibfield  {author} {\bibinfo {author} {\bibfnamefont {M.}~\bibnamefont
  {Namburi}}, \bibinfo {author} {\bibfnamefont {S.}~\bibnamefont
  {Krithivasan}}, \ and\ \bibinfo {author} {\bibfnamefont {S}~\bibnamefont
  {Ansumali}},\ }\bibfield  {title} {\enquote {\bibinfo {title}
  {Crystallographic lattice {B}oltzmann method},}\ }\href@noop {} {\bibfield
  {journal} {\bibinfo  {journal} {Sci. Rep.}\ }\textbf {\bibinfo {volume}
  {6}},\ \bibinfo {pages} {27172} (\bibinfo {year} {2016})}\BibitemShut
  {NoStop}%
\bibitem [{\citenamefont {Frenkel}\ and\ \citenamefont {Smit}(2001)}]{frenkel}%
  \BibitemOpen
  \bibfield  {author} {\bibinfo {author} {\bibfnamefont {D.}~\bibnamefont
  {Frenkel}}\ and\ \bibinfo {author} {\bibfnamefont {B.}~\bibnamefont {Smit}},\
  }\href@noop {} {\emph {\bibinfo {title} {Understanding Molecular Simulation:
  From Algorithms to Applications}}}\ (\bibinfo  {publisher} {Academic Press,
  2001},\ \bibinfo {year} {2001})\BibitemShut {NoStop}%
\bibitem [{\citenamefont {Kadanoff}\ and\ \citenamefont
  {Baym}(1962)}]{kadanoff}%
  \BibitemOpen
  \bibfield  {author} {\bibinfo {author} {\bibfnamefont {L.~P.}\ \bibnamefont
  {Kadanoff}}\ and\ \bibinfo {author} {\bibfnamefont {G.}~\bibnamefont
  {Baym}},\ }\href@noop {} {\emph {\bibinfo {title} {Quantum Statistical
  Mechanics: Green's Function Methods in Equilibrium and Nonequilibrium
  Problems}}}\ (\bibinfo  {publisher} {W.A. Benjamin},\ \bibinfo {year}
  {1962})\BibitemShut {NoStop}%
\bibitem [{\citenamefont {Rammer}(2007)}]{rammer}%
  \BibitemOpen
  \bibfield  {author} {\bibinfo {author} {\bibfnamefont {J.}~\bibnamefont
  {Rammer}},\ }\href@noop {} {\emph {\bibinfo {title} {Quantum Field Theory of
  Non-Equilibrium States}}}\ (\bibinfo  {publisher} {Cambridge University
  Press},\ \bibinfo {year} {2007})\BibitemShut {NoStop}%
\bibitem [{\citenamefont {Hillery}\ \emph {et~al.}(1984)\citenamefont
  {Hillery}, \citenamefont {O'Connell}, \citenamefont {Scully},\ and\
  \citenamefont {Wigner}}]{wig_physrep}%
  \BibitemOpen
  \bibfield  {author} {\bibinfo {author} {\bibfnamefont {M.}~\bibnamefont
  {Hillery}}, \bibinfo {author} {\bibfnamefont {R.~F.}\ \bibnamefont
  {O'Connell}}, \bibinfo {author} {\bibfnamefont {M.~O.}\ \bibnamefont
  {Scully}}, \ and\ \bibinfo {author} {\bibfnamefont {E.~P.}\ \bibnamefont
  {Wigner}},\ }\bibfield  {title} {\enquote {\bibinfo {title} {Distribution
  functions in physics: {F}undamentals},}\ }\href@noop {} {\bibfield  {journal}
  {\bibinfo  {journal} {Physics Reports}\ }\textbf {\bibinfo {volume} {106}},\
  \bibinfo {pages} {121--167} (\bibinfo {year} {1984})}\BibitemShut {NoStop}%
\bibitem [{\citenamefont {Sol\'orzano}\ \emph {et~al.}(2018)\citenamefont
  {Sol\'orzano}, \citenamefont {Mendoza}, \citenamefont {Succi},\ and\
  \citenamefont {Herrmann}}]{lat_wig}%
  \BibitemOpen
  \bibfield  {author} {\bibinfo {author} {\bibfnamefont {S.}~\bibnamefont
  {Sol\'orzano}}, \bibinfo {author} {\bibfnamefont {M.}~\bibnamefont
  {Mendoza}}, \bibinfo {author} {\bibfnamefont {S.}~\bibnamefont {Succi}}, \
  and\ \bibinfo {author} {\bibfnamefont {H.~J.}\ \bibnamefont {Herrmann}},\
  }\bibfield  {title} {\enquote {\bibinfo {title} {Lattice {W}igner
  equation},}\ }\href@noop {} {\bibfield  {journal} {\bibinfo  {journal} {Phys.
  Rev. E}\ }\textbf {\bibinfo {volume} {97}},\ \bibinfo {pages} {013308}
  (\bibinfo {year} {2018})}\BibitemShut {NoStop}%
\bibitem [{\citenamefont {Bhatnagar}\ \emph {et~al.}(1954)\citenamefont
  {Bhatnagar}, \citenamefont {Gross},\ and\ \citenamefont {Krook}}]{bgk}%
  \BibitemOpen
  \bibfield  {author} {\bibinfo {author} {\bibfnamefont {P.~L.}\ \bibnamefont
  {Bhatnagar}}, \bibinfo {author} {\bibfnamefont {E.~P.}\ \bibnamefont
  {Gross}}, \ and\ \bibinfo {author} {\bibfnamefont {M.}~\bibnamefont
  {Krook}},\ }\bibfield  {title} {\enquote {\bibinfo {title} {A model for
  collision processes in gases. i. {S}mall amplitude processes in charged and
  neutral one-component systems},}\ }\href@noop {} {\bibfield  {journal}
  {\bibinfo  {journal} {Phys. Rev.}\ }\textbf {\bibinfo {volume} {94}},\
  \bibinfo {pages} {511--525} (\bibinfo {year} {1954})}\BibitemShut {NoStop}%
\bibitem [{\citenamefont {Coelho}\ \emph {et~al.}(2014)\citenamefont {Coelho},
  \citenamefont {Ilha}, \citenamefont {Doria}, \citenamefont {Pereira},\ and\
  \citenamefont {Aibe}}]{coelho2014lattice}%
  \BibitemOpen
  \bibfield  {author} {\bibinfo {author} {\bibfnamefont {R.C.V.}\ \bibnamefont
  {Coelho}}, \bibinfo {author} {\bibfnamefont {A.}~\bibnamefont {Ilha}},
  \bibinfo {author} {\bibfnamefont {M.~M.}\ \bibnamefont {Doria}}, \bibinfo
  {author} {\bibfnamefont {R.M.}\ \bibnamefont {Pereira}}, \ and\ \bibinfo
  {author} {\bibfnamefont {V.~Y.}\ \bibnamefont {Aibe}},\ }\bibfield  {title}
  {\enquote {\bibinfo {title} {Lattice {B}oltzmann method for bosons and
  fermions and the fourth-order hermite polynomial expansion},}\ }\href@noop {}
  {\bibfield  {journal} {\bibinfo  {journal} {Phys. Rev. E}\ }\textbf {\bibinfo
  {volume} {89}},\ \bibinfo {pages} {043302} (\bibinfo {year}
  {2014})}\BibitemShut {NoStop}%
\bibitem [{\citenamefont {Sbragaglia}\ \emph
  {et~al.}(2007{\natexlab{b}})\citenamefont {Sbragaglia}, \citenamefont
  {Benzi}, \citenamefont {Biferale}, \citenamefont {Succi}, \citenamefont
  {Sugiyama},\ and\ \citenamefont {Toschi}}]{sbragaglia2007generalized}%
  \BibitemOpen
  \bibfield  {author} {\bibinfo {author} {\bibfnamefont {M.}~\bibnamefont
  {Sbragaglia}}, \bibinfo {author} {\bibfnamefont {R.}~\bibnamefont {Benzi}},
  \bibinfo {author} {\bibfnamefont {L.}~\bibnamefont {Biferale}}, \bibinfo
  {author} {\bibfnamefont {S.}~\bibnamefont {Succi}}, \bibinfo {author}
  {\bibfnamefont {K.}~\bibnamefont {Sugiyama}}, \ and\ \bibinfo {author}
  {\bibfnamefont {F.}~\bibnamefont {Toschi}},\ }\bibfield  {title} {\enquote
  {\bibinfo {title} {Generalized lattice boltzmann method with multirange
  pseudopotential},}\ }\href@noop {} {\bibfield  {journal} {\bibinfo  {journal}
  {Phys. Rev. E}\ }\textbf {\bibinfo {volume} {75}},\ \bibinfo {pages} {026702}
  (\bibinfo {year} {2007}{\natexlab{b}})}\BibitemShut {NoStop}%
\bibitem [{\citenamefont {Solorzano}\ \emph {et~al.}(2018)\citenamefont
  {Solorzano}, \citenamefont {Mendoza}, \citenamefont {Succi},\ and\
  \citenamefont {Herrmann}}]{solorzano2018lattice}%
  \BibitemOpen
  \bibfield  {author} {\bibinfo {author} {\bibfnamefont {S.}~\bibnamefont
  {Solorzano}}, \bibinfo {author} {\bibfnamefont {M.}~\bibnamefont {Mendoza}},
  \bibinfo {author} {\bibfnamefont {S.}~\bibnamefont {Succi}}, \ and\ \bibinfo
  {author} {\bibfnamefont {H.~J.}\ \bibnamefont {Herrmann}},\ }\bibfield
  {title} {\enquote {\bibinfo {title} {Lattice wigner equation},}\ }\href@noop
  {} {\bibfield  {journal} {\bibinfo  {journal} {Physical Review E}\ }\textbf
  {\bibinfo {volume} {97}},\ \bibinfo {pages} {013308} (\bibinfo {year}
  {2018})}\BibitemShut {NoStop}%
\bibitem [{\citenamefont {Brewer}\ \emph {et~al.}(2016)\citenamefont {Brewer},
  \citenamefont {Mendoza}, \citenamefont {Young},\ and\ \citenamefont
  {Romatschke}}]{brewer2016lattice}%
  \BibitemOpen
  \bibfield  {author} {\bibinfo {author} {\bibfnamefont {J.}~\bibnamefont
  {Brewer}}, \bibinfo {author} {\bibfnamefont {M.}~\bibnamefont {Mendoza}},
  \bibinfo {author} {\bibfnamefont {R.~E.}\ \bibnamefont {Young}}, \ and\
  \bibinfo {author} {\bibfnamefont {P.}~\bibnamefont {Romatschke}},\ }\bibfield
   {title} {\enquote {\bibinfo {title} {Lattice boltzmann simulations of a
  strongly interacting two-dimensional fermi gas},}\ }\href@noop {} {\bibfield
  {journal} {\bibinfo  {journal} {Physical Review A}\ }\textbf {\bibinfo
  {volume} {93}},\ \bibinfo {pages} {013618} (\bibinfo {year}
  {2016})}\BibitemShut {NoStop}%
\bibitem [{\citenamefont {Succi}\ and\ \citenamefont
  {Montessori}(2024)}]{succi2024keldysh}%
  \BibitemOpen
  \bibfield  {author} {\bibinfo {author} {\bibfnamefont {S.}~\bibnamefont
  {Succi}}\ and\ \bibinfo {author} {\bibfnamefont {A.}~\bibnamefont
  {Montessori}},\ }\bibfield  {title} {\enquote {\bibinfo {title}
  {Keldysh-lattice {B}oltzmann approach to quantum nanofluidics},}\ }\href@noop
  {} {\bibfield  {journal} {\bibinfo  {journal} {arXiv preprint
  arXiv:2403.15768}\ } (\bibinfo {year} {2024})}\BibitemShut {NoStop}%
\bibitem [{\citenamefont {Kovtun}\ \emph {et~al.}(2005)\citenamefont {Kovtun},
  \citenamefont {Son},\ and\ \citenamefont {Starinets}}]{kovtun}%
  \BibitemOpen
  \bibfield  {author} {\bibinfo {author} {\bibfnamefont {P.~K.}\ \bibnamefont
  {Kovtun}}, \bibinfo {author} {\bibfnamefont {D.~T.}\ \bibnamefont {Son}}, \
  and\ \bibinfo {author} {\bibfnamefont {A.~O.}\ \bibnamefont {Starinets}},\
  }\bibfield  {title} {\enquote {\bibinfo {title} {Viscosity in strongly
  interacting quantum field theories from black hole physics},}\ }\href@noop {}
  {\bibfield  {journal} {\bibinfo  {journal} {Phys. Rev. Lett.}\ }\textbf
  {\bibinfo {volume} {94}},\ \bibinfo {pages} {111601} (\bibinfo {year}
  {2005})}\BibitemShut {NoStop}%
\bibitem [{\citenamefont {Trachenko}\ and\ \citenamefont
  {Brazhkin}(2020)}]{trachenko}%
  \BibitemOpen
  \bibfield  {author} {\bibinfo {author} {\bibfnamefont {K.}~\bibnamefont
  {Trachenko}}\ and\ \bibinfo {author} {\bibfnamefont {V.~V.}\ \bibnamefont
  {Brazhkin}},\ }\bibfield  {title} {\enquote {\bibinfo {title} {Minimal
  quantum viscosity from fundamental physical constants},}\ }\href@noop {}
  {\bibfield  {journal} {\bibinfo  {journal} {Science Advances}\ }\textbf
  {\bibinfo {volume} {6}},\ \bibinfo {pages} {eaba3747} (\bibinfo {year}
  {2020})}\BibitemShut {NoStop}%
\bibitem [{\citenamefont {Kr\"uger}\ \emph {et~al.}(2017)\citenamefont
  {Kr\"uger}, \citenamefont {Kusumaatmaja}, \citenamefont {Kuzmin},
  \citenamefont {Shardt}, \citenamefont {Silva},\ and\ \citenamefont
  {Viggen}}]{halim}%
  \BibitemOpen
  \bibfield  {author} {\bibinfo {author} {\bibfnamefont {T.}~\bibnamefont
  {Kr\"uger}}, \bibinfo {author} {\bibfnamefont {H.}~\bibnamefont
  {Kusumaatmaja}}, \bibinfo {author} {\bibfnamefont {A.}~\bibnamefont
  {Kuzmin}}, \bibinfo {author} {\bibfnamefont {O.}~\bibnamefont {Shardt}},
  \bibinfo {author} {\bibfnamefont {G.}~\bibnamefont {Silva}}, \ and\ \bibinfo
  {author} {\bibfnamefont {E.~M.}\ \bibnamefont {Viggen}},\ }\href {\doibase
  10.1007/978-3-319-44649-3} {\emph {\bibinfo {title} {The {L}attice
  {B}oltzmann {M}ethod: {P}rinciples and {P}ractice}}}\ (\bibinfo  {publisher}
  {Springer International Publishing},\ \bibinfo {year} {2017})\BibitemShut
  {NoStop}%
\end{thebibliography}%

\end{document}